\documentclass[english,aps,prl,twocolumn,graphics,graphicx,superscriptaddress]{revtex4-1}
\usepackage[latin9]{inputenc}
\usepackage{amsmath,mathtools}
\usepackage{amssymb}
\usepackage{graphicx}
\usepackage{verbatim}
\usepackage{braket}
\usepackage{soul}
\usepackage{bm}
\usepackage{gensymb}
\usepackage{diagbox}
\usepackage{siunitx}
\usepackage[english]{babel}
\usepackage{mathtools}
\usepackage{newtxtext}  
\usepackage{amsmath}
\usepackage{amsfonts}
\usepackage{physics}
\usepackage[varbb,varg]{newtxmath} 
\usepackage{bm,dsfont,eucal}

\renewcommand\vec{\boldsymbol}

\begin{document}
\title{Landau-Level Mixing and SU(4) Symmetry Breaking in Graphene}

\author{Nemin Wei}
\affiliation{Department of Physics, Yale University, New Haven, CT 06520, USA}
\author{Guopeng Xu}
\affiliation{Department of Physics and Astronomy, University of Kentucky, Lexington, Kentucky 40506-0055, USA}
\author{Inti Sodemann Villadiego}
\affiliation{Institut fur Theoretische Physik, Universitat Leipzig, Bruderstrabe 16, 04103, Leipzig, Germany}
\author{Chunli Huang}
\affiliation{Department of Physics and Astronomy, University of Kentucky, Lexington, Kentucky 40506-0055, USA}
\date{\today} 

\begin{abstract}
Recent scanning tunneling microscopy experiments on graphene at charge neutrality under strong magnetic fields have uncovered a ground state characterized by Kekul\'e distortion (KD). In contrast, non-local spin and charge transport experiments in double-encapsulated graphene, which has a higher dielectric constant, have identified an antiferromagnetic (AF) ground state. We propose a mechanism to reconcile these conflicting observations, by showing that Landau-level mixing can drive a transition from AF to KD with the reduction of the dielectric screening. Our conclusion is drawn from studying the effect of Landau-level mixing on the lattice-scale, valley-dependent interactions to leading order in graphene's fine structure constant $\kappa = e^2/(\hbar v_F \epsilon)$. This analysis provides three key insights: 1) Valley-dependent interactions remain predominantly short-range with the $m=0$ Haldane pseudopotential being at least an order of magnitude greater than the others, affirming the validity of delta-function approximation for these interactions. 2) The phase transition between the AF and KD states is driven by the microscopic process in the double-exchange Feynman diagram. 3) The magnitudes of the coupling constants are significantly boosted by remote Landau levels. Our model also provides a theoretical basis for numerical studies of fractional quantum Hall states in graphene.

\end{abstract}
\maketitle

\textit{Introduction:}
The rich interplay between interactions, topology, and the approximate spin-valley SU(4) symmetry in graphene quantum Hall physics continues to spark puzzles even after two decades of research \cite{halperin2020fractional, alicea2006iqhe, nomura2006QHF, yang2006grapheneQHF, checkelsky2009divergent, nomura2009KT, dean2011multicomponent, goerbig2011graphene, sodemann2014fractional, spanton2018observation, zibrov2018even, zhou2019skyrmion,veyrat2020helical, farahi2023broken} and could provide insights into more complex correlated states such as those emerging in moir\'e superlattice flatbands \cite{lu2023fractional,sharpe2019emergent,serlin2020intrinsic}. A recent notable discovery is the Kekul\'e distorted (KD) state identified by scanning tunneling microscopy (STM) experiments in charge neutral monolayer graphene \cite{li2019STM,liu2022visualizing,coissard2022imaging}, which is in tension with the antiferromagnetic (AF) state supported by earlier non-local spin and charge transport experiments \cite{zhou2022strong, wei2018electrical, stepanov2018long,young2014tunable,fu2021gapless,wei2021scattering}. We present a model to reconcile these diverging observations. 


Kharitonov \cite{kharitonov2012phase} proposed that the phase diagram of neutral graphene under a strong magnetic field can be conveniently represented in a two-dimensional parameter space defined by $u_\perp$ and $u_z$.
 These phenomenological parameters, describing the strength of valley-exchange ($u_\perp$) and valley-antisymmetric ($u_z$) interactions that breaks the SU(4) symmetry in the zeroth Landau level (LL), are influenced by electron-electron and electron-phonon potentials.
Given their complex origins, they are often regarded as experimental fitting parameters \cite{zibrov2018even,zhou2022strong,liu2022visualizing}.

\begin{figure}
    \centering
    \includegraphics[width=1.0\columnwidth]{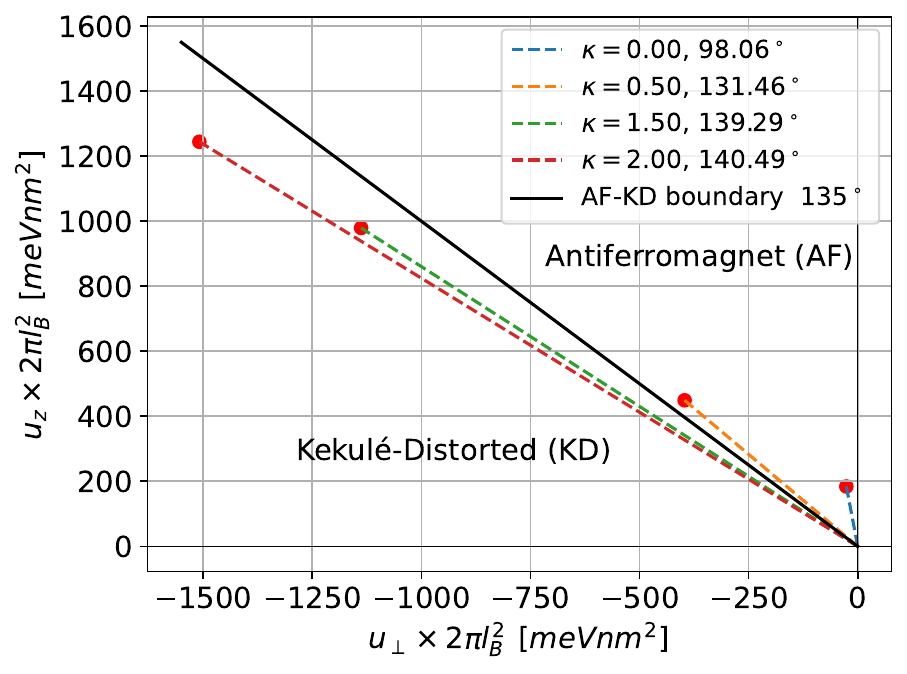}
    \caption{The flow of $u_\perp(\kappa)$ and $u_{z}(\kappa)$ with graphene's fine structure constant, $\kappa$, indicates a ground state transition from the AF phase to the KD phase as $\kappa$ increases from $0$ to $2$. The dashed lines serve as visual guides and angles in the inset are $\arctan(u_{z}/u_{\perp})$. In the absence of Zeeman splitting, the AF-KD phase boundary is marked by the black $135^\circ$ line. For comparison, we note $\Delta_Z 2\pi l_B^2=492\text{meV}\text{nm}^2$ where $\Delta_Z=2\mu_B B$ is the Zeeman energy.}
    \label{fig1}
\end{figure}

In this study, we present a comprehensive analysis of the effect of Landau level mixing on the $(u_\perp,u_z)$ to leading order of graphene's fine-structure constant, $\kappa = e^2/(\hbar v_F \epsilon)$. The constant
$\kappa$,  which depends on the electric charge $e$, Planck's constant $\hbar$, the Fermi velocity $v_F$, and the adjustable environmental dielectric constant $\epsilon$, controls the strength of the SU(4) symmetric long-range Coulomb interactions and the extent of Landau-Level mixing, as noted in previous studies \cite{peterson2014effects,sodemann2014fractional,sodemann2013landau}.
Our study reveals that as $\kappa$ increases (\textit{i.e.}, Landau level mixing becomes stronger), $u_\perp(\kappa)$ decreases more rapidly than $u_z(\kappa)$ increases. This triggers a transition from AF to a KD, as illustrated in Fig.~\ref{fig1}. 
Thus, the variation in Landau-level mixing across different samples leads to the emergence of either the KD or AF state. This trend is consistent with experimental findings: double-encapsulated samples used in transport experiments, with their larger $\epsilon$
(and consequently smaller 
$\kappa$), exhibit AF order, whereas un-encapsulated samples in STM experiments display KD order.



\textit{The Standard Model of Monolayer Graphene:}
At the atomic scale, interactions between electrons in graphene exhibit weak sublattice dependence consistent with its $C_{6v}$ symmetry. In the dilute doping regime, a specific subset of Bloch states --those with energies below an energy cutoff $\Lambda$ and concentrated at the Brillouin zone corners 
-- form the basis for constructing the long-wavelength field theory. A practical method to incorporate atomic-scale interactions into this long-wavelength theory is to model them as Dirac delta functions with corresponding bare coupling constants computed from the three dimensional atomic wavefunctions. This methodology leads to the formulation of what we refer to as "graphene's standard model", encompassing the band Hamiltonian $H_{0}$, the SU(4) symmetric long-range Coulomb interaction $H_{s}$, and the lattice-scale interactions $H_{a}$:
\cite{lemonik2012competing,aleiner2007spontaneous}
\begin{subequations}\label{eq:H}
\begin{equation} \label{eq:H0}
    H_{0} = \sum_{i=x,y}v_F\int \psi^{\dagger}(\bm r)\hat{\Pi}_{i}\sigma^{i}\psi(\bm r)\,d^2{r} 
\end{equation}
\begin{equation}
    H_{s} = \frac{e^2}{2\epsilon}\iint \frac{\psi^{\dagger}(\bm r)\psi^{\dagger}(\bm r')\psi(\bm r')\psi(\bm r)}{ |\vec{r}-\vec{r'}|}\, d^2 r d^2 r'
\end{equation}
\begin{equation}\label{eq:hasym}
    H_{a}= \frac{1}{2}\sum_{\substack{\alpha,\mu =\\0,x,y,z}}g_{\alpha\mu}^{0}\int :\left[\psi^{\dagger}(\bm r)\tau^{\alpha}\sigma^{\mu}\psi(\bm r)\right]^2: \; d^2 r
\end{equation}
\end{subequations}
Here $v_F$ is the Fermi velocity, $\hat{\Pi}_i = -i\hbar\partial_{i}+e A_{i}(\bm r)$ is the kinetic momentum operator with the vector potential $A_i$, $\epsilon$ is dielectric constant, and the basis  $\psi=(\psi_{\uparrow},\psi_{\downarrow})^{T}$ with $\psi_{s}=(\psi_{KAs},\psi_{KBs},\psi_{K'Bs},-\psi_{K'As})^{T}$ where $s$ can be $\uparrow$ or $\downarrow$. $\tau^{\alpha},\sigma^{\mu}$ are Pauli matrices operating in the valley and sublattice subspace, respectively. 

Eq.~\eqref{eq:hasym} contains 16 valley-and sublattice-dependent interactions. However, under $C_{6v}$ and lattice translation symmetry, $g_{\alpha x}^{0}=g_{\alpha y}^{0}\equiv g_{\alpha \perp}^{0}$, $g_{x\mu}^{0}=g_{y\mu}^{0}\equiv g_{\perp\mu}^{0}$ and we neglect $g_{00}^{0}$ since it is dwarfed by long-range Coulomb interaction. Thus, Eq.~\eqref{eq:hasym} is characterized by 8 independent parameters or coupling-constants\cite{aleiner2007spontaneous}: \begin{equation}
g_{0z}^{0}\,,\,g_{0\perp}^{0}\,,\,g_{z0}^{0}\,,\,g_{z\perp}^{0}\,,\,g_{z z}^{0}\,,\,g_{\perp 0}^{0}\,,\,g_{\perp \perp}^{0}\,,\,g_{\perp z}^{0}.
\end{equation}
The coupling constants $g_{\alpha\mu}^0$
  are influenced by electron-electron repulsion and electron-phonon interactions, with the superscript 
$0$
denoting their role as `bare' inputs in the theory with cutoff $\Lambda$. A key finding of our work is that Landau-level mixing significantly renormalizes these atomic-scale interactions when reducing the energy from $\Lambda$ to the cyclotron energy.


In the Supplementary Material, we provide detailed calculations of estimates for the bare coupling constants, based on the assumption that the Bloch functions are linear combinations of (3D) atomic  $p_z$ orbitals. It is important to note that convergence in real space, when summing atomic orbitals, can be notably slow and may lead to inaccuracies \cite{knothe2020quartet, supmat}. 
To circumvent these issues, performing these calculations in momentum space has proven more effective. Our calculation suggests that the intra-sublattice inter-valley-exchange interaction $g_{\perp\perp}^{0}$ is the dominant lattice-scale interaction, in particular $g_{\perp\perp}^{0}>g_{zz}^{0}$ \cite{supmat, raines2021spin}. Upon incorporating electron-phonon contributions, we obtain the following estimates (constants not listed below are found to be negligible):
\begin{equation}\label{eq:g_estimate}
(g_{\perp \perp}^0,g_{zz}^0 , g_{\perp z}^{0},\, g_{z\perp}^{0})
= (269 ,  184 , -26, -27)\text{meV}\cdot\text{nm}^2.
\end{equation}


\begin{figure}
    \centering
    \includegraphics[width=0.9\linewidth]{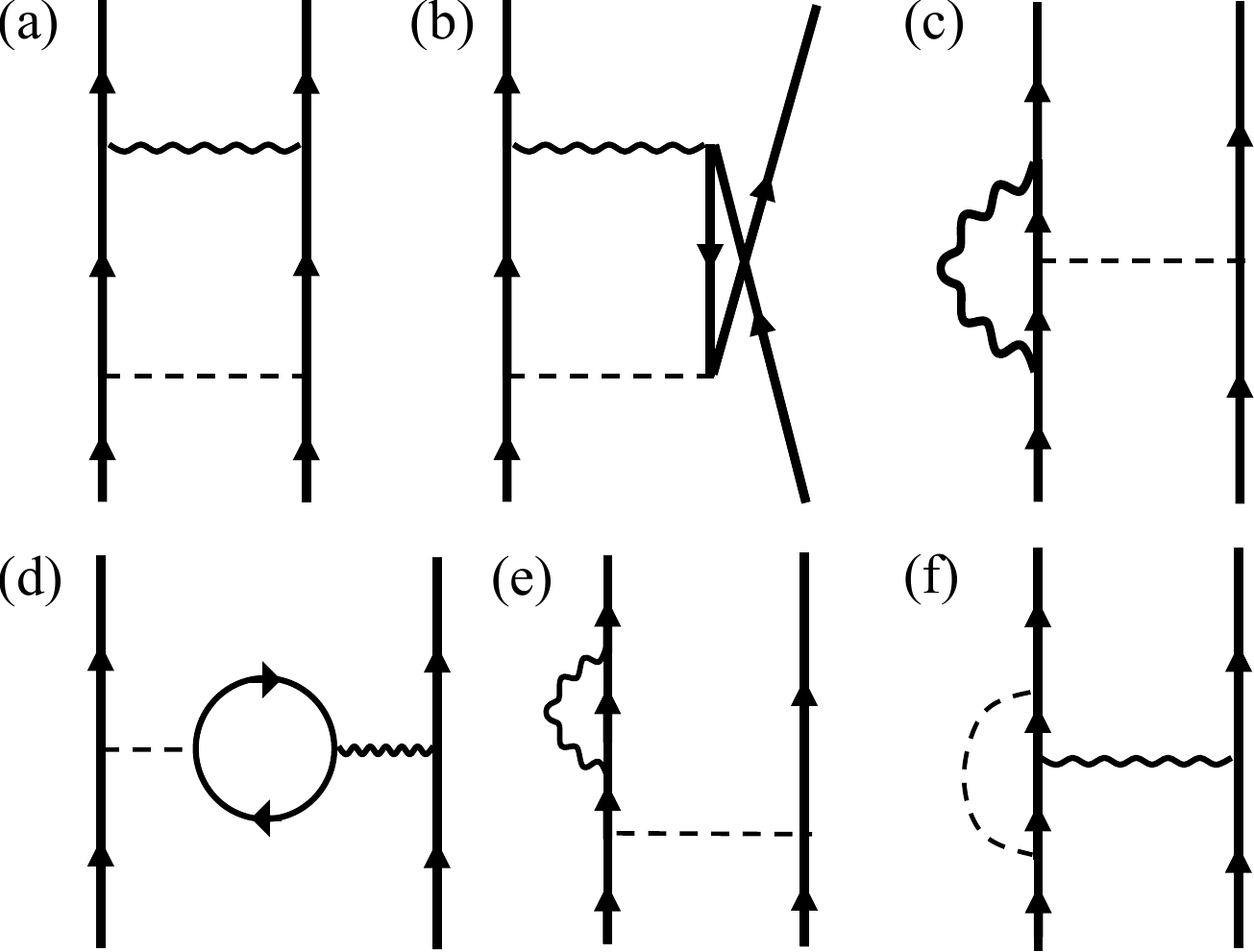}
    \caption{(a)-(c) Feynman diagrams representing the $\mathcal{O}(\kappa)$ corrections to two-body valley-dependent Haldane pseudopotentials. Solid lines are fermionic propagators and the external lines are in the $n=0$ Landau level. The curly and dashed lines stand for the long-range SU(4) Coulomb and the contact sublattice-and valley-dependent interactions, respectively. Swapping the curly and dashed lines in (a) and (b) does not change the corrections to Haldane pseudopotentials and for (c) either vertex of the dashed line can be dressed. (d)-(f) Examples of diagrams that do not contribute to the valley anisotropic interactions. (d) and (e) vanish. (f) yields SU(4) symmetric corrections and thus can be neglected.}
    \label{fig:feynman_diagram}
\end{figure}

\textit{Landau-level mixing and the modifications of valley-dependent interactions:}
%
In a constant magnetic field (i.e.~$\nabla \times \vec{A}=-B\hat{z}$ in Eq.\eqref{eq:H0}), the eigenstates of $H_0$ for spin-valley flavor are denoted as $\ket{\overline{\xi n}, m}$, where $\overline{\xi n}$ represents the Landau levels $n=0,1,2...$,  $\xi=\pm 1$, and $m$ enumerates the guiding center degeneracy of each Landau level. The wavefunction is a sublattice spinor,
\begin{equation}\label{eq:wavefunction}
    \ket{\overline{\xi n}}=(\sqrt{2})^{\delta_{n,0}-1}\begin{pmatrix}
    \ket{n}\\
    (1-\delta_{n,0})\xi\ket{n-1}
    \end{pmatrix},
\end{equation}
where each state has a kinetic energy of $\epsilon_{\overline{\xi n}}=\xi\hbar v_{F}\sqrt{2n}/l_{B}$.
The quantum state $\ket{n}\equiv a^{\dagger n}\ket{0}/\sqrt{n!}$ and $\ket{m}\equiv b^{\dagger m}\ket{0}/\sqrt{m!}$ are constructed from two sets of ladder operators constructed from kinematic momentum and guiding centers: $a^{\dagger}=(\hat{\Pi}_{x}-i\hat{\Pi}_y)l_B/\sqrt{2}$ and $b^{\dagger}=(\hat{\eta}_{x}+i\hat{\eta}_y)/\sqrt{2}l_B$ where $\hat{\bm\eta} = \bm r - l_B^2\bm z\times\hat{\bm \Pi}$. In the symmetric gauge, we can loosely call $m$ the angular momentum although the physical angular momentum is $m-n$.

Since any translationally and rotationally invariant two-body interaction Hamiltonian $\hat{V}$ can be decomposed into the Haldane pseudopotentials using the center of mass and relative coordinates $|N M\rangle_{c}|n m\rangle_{r}$  \cite{haldane1983fractional},
\begin{align}\label{eq:pseudopotential}
    &{}_c\bra{N'M'}{}_r\bra{n'm'}\hat{V}\ket{NM}_{c}\ket{nm}_r \notag\\
    &={}_{r}\bra{n'm'}\hat{V}\ket{nm}_{r}\delta_{m'-n',m-n}\delta_{N',N}\delta_{M',M}.
\end{align}
For two interacting electrons in the zeroth LL of our interest, $n=n'=0$. In the spirit of Larmor's theorem, we can eliminate the Zeeman field from the equation of motion in a rotating frame and recover the spin SU(2) symmetry. Thus, the effective interactions can be represented in a spin-independent form. Moreover, the $C_{6v}$, lattice momentum conservation, and the valley-sublattice locking of the zeroth LL, enforce effectively an U(1) valley conservation, leading to the Haldane pseudopotentials $\prescript{}{r}{\bra{0m}}V\ket{0m}_{r} = V_0(m) + V_{a}(m)$, where $V_0(m)$ is spin-valley SU(4) Haldane pseudopotential that has been studied previously \cite{peterson2013more} and
\begin{equation} \label{eq:pseudopotential_zLL}
    \hat{V}_{a}(m) = \frac{u_{z}(m)}{2}
    \tau_{i}^{z}\tau_{j}^{z} + \frac{u_{\perp}(m)}{2}(\tau_{i}^{x}\tau_{j}^{x} + \tau_{i}^{y}\tau_{j}^{y}),
\end{equation}
are the valley dependent pseudopotentials arising from $H_{sv}$ interactions in Eq.~\eqref{eq:hasym}. At the zeroth order in interactions, the effective Hamiltonian is $\hat{V}^{(1)}=\mathcal{P}(\hat{H}_{s}+\hat{H}_{a})\mathcal{P}$ with $\mathcal{P}$ being the projector into the degenerate non-interacting ground-state manifold where $n<0$ LLs are fully occupied and $n>0$ LLs empty. Because a pair of electrons of a nonzero relative angular momentum do not experience contact interactions \cite{supmat}, $u_{\alpha}^{(1)}(m)=u_{\alpha}\delta_{m,0}$, where $u_{\alpha}= (g_{\alpha 0}^{0}+g_{\alpha z}^{0})/2\pi l_{B}^2$ is the notation used in Ref.~\cite{kharitonov2012phase}. However, Landau level mixing generates perturbative corrections to these Haldane pseudotentials and one of our main goals is to compute such corrections.

The leading order correction to $\hat{V}^{(1)}$ can be derived from second-order perturbation theory \cite{sodemann2013landau}. Our focus here is on the effect of dielectric screening on valley-dependent interactions, which is captured by the following Hamiltonian, comprising one $\hat{H}_s$ and one $\hat{H}_a$:
\begin{align} \label{eq:effective_a}
    \hat{V}_{sa}^{(2)} &= -\mathcal{P}\hat{H}_{s} \mathcal{P}_{\perp}\frac{1}{\hat{H}_0 - E_{0}} \mathcal{P}_{\perp}\hat{H}_{a}\mathcal{P} + h.c. 
\end{align}
 Here $E_0$ is the energy of degenerate noninteracting ground-states and $\mathcal{P}_{\perp}=\mathcal{I}-\mathcal{P}$. 
Because three-body interactions are forbidden in the zeroth LL of monolayer graphene by the particle-hole symmetry \cite{peterson2013more} and the one-body self-energy generated by $\hat{V}_{sa}^{(2)}$ is SU(4) symmetric, the only nontrivial terms in Eq.~\eqref{eq:effective_a} are the two-body interactions which can be delineated into Feynman diagrams in Fig.~\ref{fig:feynman_diagram} \cite{supmat}.
We call the three non-vanishing Feynman diagrams (a) the particle-particle channel, (b) the double-exchange channel, and (c) the vertex correction. Their contributions are parametrized as follows:
\begin{equation}\label{eq:u2}
u_{\alpha}^{(2)}(m)=\frac{\kappa}{2\pi l_{B}^2}\sum_{\mu=0,x,y,z}(c_{m,a}^{\mu}+c_{m,b}^{\mu}+c_{m,c}^{\mu}) g_{\alpha\mu}^{0}    
\end{equation}
Because of our choice of pseudospin components (see text below Eqs.~\eqref{eq:H}), both valleys sharing identical orbital wave functions as per Eq.~\eqref{eq:wavefunction}, the coefficients $c_{m,a/b/c}^{\mu}$ are independent of the valley index $\alpha$.

Evaluating the Feynman diagram in Fig.~\ref{fig:feynman_diagram}a, we find that the valley-sublattice locking unique to the zeroth LL of graphene, $\sigma^{z}\ket{\overline{0}m}=\ket{\overline{0}m}$, leads to the vanishing interacting matrix elements $\bra{\bar{n}_i,\bar{n}_j}\sigma_i^{+}\sigma_j^{-}\ket{\bar{0}_i,\bar{0}_j}=0$, and $c_{m,a}^{\perp}\equiv c_{m,a}^{x}+c_{m,a}^{y}=0$. Additionally, due to the contact valley-dependent interaction, the particle-particle ladder diagram only induces corrections to $u_{\alpha}^{(2)}(m=0)$, with \cite{supmat}
\begin{align} 
    \frac{c_{m=0,a}^{0,z}}{4\pi}= &-2\sum_{n=1}^{2N_{c}}\prescript{}{r}{\bra{00}}\delta(\bm r_{ij})\ket{n n}_{r}\prescript{}{r}{\bra{n n}}r_{ij}^{-1}\ket{00}_{r} \notag\\
    &\quad\times \sum_{n_{i,j}=-\infty}^{\infty}\frac{\Theta(n_in_j)}{|\epsilon_{\overline{n_i}}+\epsilon_{\overline{n_j}}|}|\prescript{}{c}\langle 0|\prescript{}{r}\langle n||n_{i}|,|n_{j}|\rangle|^2, \notag\\
    =&\sum_{n=1}^{2N_{c}}\frac{\Gamma(n+\frac{1}{2})}{2^{n+1} n!}\sum_{n_1=0}^{n}
   \frac{\binom{n}{n_{1}}}{\sqrt{2n_1}+\sqrt{2(n-n_1)}}. \label{eq:cm_pp}
\end{align}
Here, $\Theta(x)$ is the Heaviside function with ${\Theta(0)=1}$ and we measure the length in the unit of $l_B$ an the kinetic energy in the unit of $\hbar v_F/l_B$ to shorten the expressions.
Asymptotically, $\Gamma(n+\frac{1}{2})/n!\sim n^{-\frac{1}{2}}$ and therefore $c_{0,a}^{0,z}$ is logarithmically divergent with respect to the LL cutoff $N_c$, which is expected from the renormalization group analysis \cite{kharitonov2012phase}.
%
%

The Feynman diagrams in Fig.~\ref{fig:feynman_diagram}b and c involve particle-hole excitations in the intermediate states and need to be calculated by a different method. We follow Ref.~\cite{sodemann2013landau,haldane1987quantum} and express the bare two-body interactions as Fourier integrals, $r_{ij}^{-1}=\int d\bm{q} e^{i\bm{q}\cdot\bm{r}_{ij}}/2\pi q$ and $\delta(\bm{r}_{ij})=\int d\bm{q}' e^{i\bm{q}'\cdot\bm{r}_{ij}}/(2\pi)^2$. For each $(\bm{q},\bm{q}')$, the scattering amplitude is diagonalized in the relative angular momentum basis $\{\ket{m}_r\}$ \cite{supmat}. The Haldane pseudopotentials can therefore be expressed as an integral,
with
\begin{align}
    \frac{c_{m,b}^{\mu}}{4\pi} = \iint_{\bm{q},\bm{q}'}&\sum_{n,n'=-\infty}^{\infty}\frac{\Theta(n)-\Theta(n')}{\epsilon_{\overline{n}}-\epsilon_{\overline{n'}}}\mathcal{V}_{n0,0n'}^{\mu}(\bm{q}')\mathcal{V}_{0n',n0}(\bm{q})\notag\\
    &\times \prescript{}{r}{\langle m|}e^{i(\bm{q}-\bm{q}')\cdot\bm{\eta}_{ij}}|m\rangle_{r} + c.c., \label{eq:cm_ph} \\
    \frac{c_{m,c}^{\mu}}{4\pi} = \iint_{\bm{q},\bm{q}'}&\sum_{n,n'=-\infty}^{\infty}\frac{\Theta(n')-\Theta(n)}{\epsilon_{\overline{n}}-\epsilon_{\overline{n'}}}\mathcal{V}_{0n,0n'}^{\mu}(\bm{q}')\mathcal{V}_{0n',0n}(\bm{q})\notag\\
    & \times e^{i\bm q'\wedge \bm{q}}\prescript{}{r}{\langle m|}e^{-i\bm{q}'\cdot\bm{\eta}_{ij}}|m\rangle_{r} + c.c.. \label{eq:cm_gamma}
\end{align}
Here $\int_{\bm q}=\int d^2q /(2\pi)^2$ and the matrix elements for the long-range Coulomb interaction $\mathcal{V}_{nn',n_1n_1'}(\bm q) = \frac{2\pi}{q}F_{n,n_1}^{0}(\bm q)F_{n',n_1'}^{0}(-\bm q)$ and the bare sublattice-and valley-dependent interactions $\mathcal{V}_{nn',n_1n_1'}^{\mu}(\bm q) = F_{n,n_1}^{\mu}(\bm q)F_{n',n_1'}^{\mu}(-\bm q)$ with the 
form factors $F_{n',n}^{\mu}(\bm q)=\bra{\overline{n'}}\sigma^{\mu} e^{i\bm q\cdot(\hat{z}\times\bm\Pi) }\ket{\overline{n}}$ for $\mu=0,x,y,z$.
%
Due to the valley-sublattice locking in the zeroth LL, $c_{m,c}^{\perp}=0$ as in the particle-particle channel.
 
\begin{table}
\caption{Coefficients of the first-order-in-$\kappa$ corrections to the valley-dependent interactions receive contributions from three non-vanishing Feynman diagrams in Fig.~\ref{fig:feynman_diagram}a (p-p), b (DE), and c(vertex), see Eq.~\eqref{eq:u2}. (a) $c_{m=0,a/b/c}^{\mu}$ are listed in this table. (b) $c_{m}^{\mu}=c_{m,a}^{\mu}+c_{m,b}^{\mu}+c_{m,c}^{\mu}$ for $m\leq 5$. The numbers in red depend logarithmically on the cutoff $N_c$ of the Landau levels. Here, $N_c=160$.}
\begin{tabular}{|c|c|c|c|}
\multicolumn{4}{@{}l}{(a) $c_{m=0,a}^{\mu},\ \ \ c_{m=0,b}^{\mu},\ \ \ c_{m=0,c}^{\mu}$}\\
\hline $\mu$ & p-p & DE & vertex \\
\hline $0$&\textcolor{red}{$-1.6773$} & \textcolor{red}{$1.5609$} & $0.3915$ \\
\hline $z$&\textcolor{red}{$-1.6773$} & \textcolor{red}{$1.5609$} & \textcolor{red}{$2.7386$} \\
\hline $\perp$&0 & \textcolor{red}{$-2.5900$} & 0\\
\hline
\end{tabular} \label{tab:m0}\\
\begin{tabular}{|c|c|c|c|c|c|c|}
\multicolumn{7}{@{}l}{(b)}\\
\hline \diagbox[]{$\mu$}{m} &0 & 1 & 2 & 3 & 4 & 5 \\
\hline $0$&$0.2799$ & $-0.1094$ & $-0.1283$ & $-0.0144$ & $-0.0173$ & $-0.0051$  \\
\hline $z$&\textcolor{red}{$2.6223$} & $0.1128$ & $-0.0627$ & $0.0149$ & $-0.0011$ & $0.0051$  \\
\hline $\perp$&\textcolor{red}{$-2.5900$} & $0.1064$ & $-0.0141$ & $0.0032$ & $-0.0001$ & $0.0006$  \\
\hline
\end{tabular}\label{tab:cm}
\end{table}

\textit{Discussion of main results:--}
The main results of this work, encapsulated in Eq.~\eqref{eq:cm_pp}, \eqref{eq:cm_ph}, and \eqref{eq:cm_gamma}, required numerical computations and they lead to the following three important conclusions:

1) Table~\ref{tab:m0}(a) shows the contributions of three Feynman diagrams to the $m=0$ component of $c_m^{\mu} (c_{m}^{\perp} \equiv c_{m}^{x}+c_{m}^{y})$. Notably, the inter-valley channel $c_{0}^{\perp}$, which is entirely derived from the double-exchange diagram, is found to be attractive ($c_{0}^{\perp}<0$). This observation is significant because it implies that a repulsive valley-and sublattice-exchange scattering $g_{\perp\perp}^{0}>0$ can  induce a negative valley-exchange Haldane pseudopotential $V_{\perp}$ in the zeroth LL. 
This contrasts sharply with conventional fermions with parabolic dispersion, where repeated exchange scattering enhances opposite-spin repulsion and suppresses spin-singlet superconductivity near a ferromagnetic critical point \cite{berk1966effect}.
In graphene, we found that opposite-valley scattering becomes attractive due to the opposite chirality of electron and hole wavefunctions \footnote{While our conclusion is based on double-exchange diagram, the Berk-Schrieffer diagram consists of an infinite series of repeated exchange diagrams}. In the context of Landau levels, the second sublattice components of wavefunctions (Eq.\eqref{eq:wavefunction}) for intermediate electron ($\xi>0$) and hole ($\xi<0$) have opposite signs, leading to an attractive interaction matrix elements $\mathcal{V}_{n0,0n'}^{\perp}<0$.

2) The red coefficients in Table I(a) highlight that these values logarithmically increase with the increasing number of Landau levels, a consequence of the linearly increasing density-of-states in our low-energy theory. This is an important effect as they amplify the magnitude of lattice-scale interactions, making them 3 to 5 times stronger than the Zeeman energy, consistent with experimental observations \cite{supmat, zibrov2018even}. 

3) Table~\ref{tab:cm}(b) details the values of $u_{\alpha}^{(2)}(m)$ for $m=0,...,5$, incorporating contributions from the aformentioned Feynman diagrams. Notably,  the $m=0$ component is significantly larger, by at least an order of magnitude, than the $m\neq0$ component. This indicates that even when contact interactions are dressed by long-range Coulomb interactions in second-order perturbation theory, the resulting interaction remains primarily short-ranged. While this is evident in the particle-particle channel ($c_{m\neq 0 ,a}^{\mu} = 0$) the particle-hole channels require further explanation.
In the particle-hole channel, the bubble diagram contribution is exactly zero leaving the exchange scattering diagrams as the primary contributors.
The bubble diagram vanishes because
the susceptibility of valley or sublattice polarization induced by charge perturbation vanishes under $C_{6v}$ symmetry. The exchange diagrams fail to generate long-range interactions, as they require substantial spatial overlap of the wavefunctions. 
This result suggests that the coexistence phase of the KD and AF states, as proposed in Ref.~\onlinecite{das2022coexistence,de2023global}, does exist, but it occupies a very small sliver of the phase diagram near the boundary of KD and AF \cite{stefanidis2023spin,stefanidis2023competing}. 




Focusing on the $m=0$ component, we obtain the final expression for the corrections to
the lattice-scale interactions to leading order in $\kappa$ are given by:
\begin{align}
    u_{\perp}(\kappa)&\equiv u_\perp+u_\perp^{(2)}(0)= \frac{
     g_{\perp z}^0\left(1+\kappa c^z_0 \right)-\kappa |c^{\perp}_0| g_{\perp \perp}^0}{2\pi l_{B}^2}, \\
    u_{z}(\kappa) &\equiv u_z+ u_z^{(2)}(0)=\frac{ 
     g_{zz}^0\left(1+\kappa c^z_0 \right)-\kappa |c^{\perp}_0| g_{z\perp}^0 }{2\pi l_{B}^2}.
\end{align}
By incorporating the estimates Eq.~\eqref{eq:g_estimate}, we derive the main results shown in Fig.~1. At $\kappa=0$ (\textit{i.e.,}~without Landau-level mixing), $u_\perp$ and $u_z$ are determined by attractive $K$-optical-phonon potential ($g_{\perp z}^{0}$) and repulsive sublattice asymmetry electron-electron repulsion ($g_{zz}^{0}$) respectively, with the latter being more dominant, leading to
$u_z>-u_{\perp}$ and thus an AF ground state.
At finite $\kappa$, the respective contributions of $g_{\perp z}^{0}$ and $g_{zz}^{0}$ to $u_\perp$ an $u_z$ are enhanced by the same factor $1+\kappa c_0^z$. However, the double-exchange Feynman diagram introduces into $u_\perp$ the on-site valley-exchange scattering ($g_{\perp \perp}^{0}$), the strongest type of bare lattice-scale interactions in monolayer graphene \cite{supmat, raines2021spin}, making $u_\perp$ decrease faster than the increases in $u_z$ and ultimately tipping the balance to $u_z<-u_{\perp}$ and the ground-state transition to the KD state, as shown in Fig.~1.

%

\textit{Conclusions and outlook:}
In summary, our study reveals a transition from antiferromagnetic  to Kekul\'e distorted states in graphene as a function of the fine-structure constant $\kappa$ occurring around $\kappa\sim0.5$. 
This transition is driven by the enhancement of Landau-level mixing in the double-exchange Feynman diagram. This diagram uniquely produces an attractive $u_\perp$ from repulsive valley-exchange scattering $g_{\perp \perp}$ due to matrix-element effects.
The logarithmic divergence of $u_{z}$ and $u_{\perp}$ agrees qualitatively with prior renormalization group analyses conducted without a magnetic field \cite{kharitonov2012phase, aleiner2007spontaneous,supmat}. This correspondence suggests that the notable enhancement of lattice-scale interactions is largely attributed to the influence of remote Landau levels \cite{supmat}. 
Our research advances the understanding of the phase diagram of charge neutral graphene under the strong magnetic field. The calculated valley-dependent Haldane pseudopotentials provide a foundation for accurate numerical studies of SU(4) symmetry breaking in graphene's fractional quantum Hall states \cite{peterson2013more}. Furthermore, our study sheds light on the complex interplay between dielectric screening, valley- and sublattice-dependent interactions, and unveils a novel mechanism for generating attractive valley-exchange interactions from bare lattice-scale electron-electron repulsion. This mechanism could be relevant in other contexts, such as superconductivity in strongly-correlated multilayer graphene systems \cite{chatterjee2022inter,qin2023functional,huang2022pseudospin}.



\textit{Acknowledgement}
We thank Jeanie Lau, J.I.A Li, X.~Liu, A.~H.~MacDonald, G.~Murthy, A. Yazdani, A.~Young, and H.~Zhou for discussions. N.~W. acknowledges the hospitality of University of Kentucky where part of this work has been performed. We acknowledge support from the Deutsche Forschungsgemeinschaft (DFG) through research grant project number 518372354.

\bibliography{references}

\clearpage
\widetext

\begin{center}
\textbf{\large Supplementary Material: Landau-Level Mixing and SU(4) Symmetry Breaking in Graphene}
\end{center}

\setcounter{equation}{0}
\setcounter{figure}{0}
\setcounter{table}{0}
\setcounter{page}{1}
\makeatletter
\renewcommand{\theequation}{S\arabic{equation}}
\renewcommand{\thefigure}{S\arabic{figure}}

\newcounter{proplabel}


\section{Estimation of the bare g-values from tight-binding calculation}

To compute the contribution of electron-electron repulsion to $g_{\alpha\mu}^{0}$,
we adopt a basic tight-binding model, assuming that the Bloch function of graphene is a linear combination of the $p_z$ atomic orbitals. For this calculation, we only require two Bloch functions situated at the non-equivalent corners, $\vec{\tau}=\pm\vec{K}$, of the Brillouin zone and their two sublattice projections. They can be grouped into a vector,
\begin{equation} 
v=\left[ u_{KA}, u_{KB}, u_{K'B},- u_{K'A}\right]^T,
\end{equation}
\begin{equation} \label{eq:bloch}    \langle\vec{r}\ket{u_{\tau\sigma}}=
\sqrt{\frac{A_0}{N}}
\sum_{\vec{R}}e^{i\vec{\tau}\cdot (\vec{R}_\sigma+\vec{R})}\phi(\vec{r}-\vec{R}_\sigma-\vec{R}).
\end{equation}
Here $\vec{R}_{\sigma}$ represents the sublattice vector so that $\vec{R}_{A}=0$ and $\vec{R}_{B}=\vec{R}_{AB}$. In our coordinate system, the origin coincides with sublattice $A$ and and $\vec{R}_{AB}$ is used to denote the relative displacement between the two carbon sublattices inside the unit cell. Consequently, the summation over
 $\vec{R}_A$ covers the set: $\{\vec{R}_A| m\vec{a}_1+n\vec{a}_2\,\forall m,n \in \mathbb{Z}$\} and 
 the summation over
 $\vec{R}_B$ covers the set: $\{\vec{R}_B| \vec{R}_{AB}+m\vec{a}_1+n\vec{a}_2\,\forall m,n \in \mathbb{Z}$\}
 where $\vec{a}_1$ and $\vec{a}_2$ are the primitive lattice vectors. 
 
 The normalized wavefunction for $p_z$ orbitals is given by 
 \begin{equation}
\phi(r,\theta,\phi)=\frac{2\tilde{Z}^2\sqrt{\tilde{Z}}}{\sqrt{96a_B^5}}re^{-\frac{\tilde{Z}r}{2a_B}}Y_1^0(\theta,\phi),
 \end{equation}
 where $\int |\phi(\vec{r})|^2 d^3r=1$.
 Here, $a_B$ is the Bohr radius and $\tilde{Z}$ is the effective nuclear charge of carbon atom. With this basis in place, $g_{\alpha\mu}^{0}$ can be directly computed as,
\begin{align}\label{eq:g0}
    g_{\alpha\mu}^{0}&=\frac{1}{4A_0}
    \sum_{ijkl} V_{ik,jl} (\tau^{\alpha}\sigma^{\mu})_{ij}   (\tau^{\alpha}\sigma^{\mu})_{kl},
\end{align}
 where the Coulomb matrix elements are described by the six-dimensional integrals,
 
 \begin{equation} \label{eq:Vc_mat_ele}
    V_{ik,jl} =e^2 \int \int_{u.c.} \frac{ v^{*}_i(\bm r_1)v_j(\bm r_1)
    v^{*}_k(\bm r_2)v_l(\bm r_2)}{|\bm r_1-\bm r_2|}\;d^3 r_1d^3r_2.
 \end{equation}
Here $A_0$ is the area of the graphene unit cell.  Next, we express Eq.~\eqref{eq:g0} as the Coulomb potential energy between overlap charges:
\begin{align} \label{eq:g0_v1}
    g_{zz} & = \frac{e^2}{4A_0}\sum _{\tau,\sigma,\tau^\prime,\sigma^\prime}sgn(\sigma)sgn(\sigma^\prime)\int\int_{u.c.} d^3\vec{r}_1\int d^3\vec{r}_2\frac{\rho_{\tau\sigma,\tau\sigma}(\vec{r}_1)\rho_{\tau^\prime\sigma^\prime, \tau^\prime\sigma^\prime}(\vec{r}_2)}{|\vec{r}_1-\vec{r}_2|} \\
    g_{\perp \perp}& = \frac{e^2}{4A_0}\sum_{\tau \sigma}\int\int_{u.c.} d^3\vec{r}_1\int d^3\vec{r}_2\frac{\rho_{\tau\sigma,\bar{\tau}\sigma}(\vec{r}_1)\rho_{\bar{\tau}\sigma, \tau\sigma}(\vec{r}_2)}{|\vec{r}_1-\vec{r}_2|} \\
    g_{z\perp}& = \frac{e^2}{4A_0}\sum_{\tau\sigma}2\int\int_{u.c.} d^3\vec{r}_1\int d^3\vec{r}_2\frac{\rho_{\tau\bar{\sigma},\tau\sigma}(\vec{r}_1)\rho_{\tau\sigma, \tau\bar{\sigma}}(\vec{r}_2)}{|\vec{r}_1-\vec{r}_2|}\\
    g_{\perp z}&=\frac{e^2}{4A_0}\sum_{\tau\sigma}2\int\int_{u.c.} d^3\vec{r}_1\int d^3\vec{r}_2\frac{\rho_{\tau\sigma,\bar{\tau}\bar{\sigma}}(\vec{r}_1)\rho_{\bar{\tau}\bar{\sigma}, \tau\sigma}(\vec{r}_2)}{|\vec{r}_1-\vec{r}_2|}.
\end{align}
We define the overlap charge density $\rho_{ij}(\vec{r})$ as $v^{*}_{i}(\vec{r})v_{j}(\vec{r})$, where $i,j$ are indices representing $KA, KB, K'A, K'B$. Note that the overlap charge  within the same sublattice, termed as the intra-sublattice overlap charge, is significantly larger than the overlap charge between different sublattices, termed as the inter-sublattice overlap charge. Consequently, we anticipate that the magnitudes of $g^0_{zz}$ and $g^0_{\perp \perp}$ will be greater compared to those of $g^0_{z\perp}$ and $g^0_{\perp z}$. When considering the intra-sublattice overlap charge, this term can be approximated to leading order by the separation distance between two $p_z$ atomic orbitals:
\begin{align}
    \rho_{\tau_i \sigma,\tau_j \sigma}(\vec{r})\equiv v^{*}_{\tau_i \sigma}(\vec{r})v_{\tau_j\sigma}(\vec{r})&=\frac{A_0}{N}\sum_{\vec{R}_1,\vec{R}_2}e^{-i\vec{\tau}_i\cdot (\vec{R}_1+\vec{R}_{\sigma})+i\vec{\tau}_j\cdot(\vec{R}_2+\vec{R}_{\sigma})}\phi(\vec{r}-\vec{R}_1-\vec{R}_{\sigma})\phi(\vec{r}-\vec{R}_2-\vec{R}_{\sigma})\\ \notag
    &\approx A_0 \sum_{\vec{R}}e^{-i(\vec{\tau}_i-\vec{\tau}_j)\cdot(\vec{R}+\vec{R}_{\sigma})}\phi^2(\vec{r}-\vec{R}-\vec{R}_{\sigma})
\end{align}
The second line above is equivalent to selecting the $N$ combinations within the summations where $\vec{R}_1$ is equal to $\vec{R}_2$. We have numerically confirmed that contributions from subsequent $R$ shells are notably smaller, showing a reduction of at least an order of magnitude. This significant decrease is mainly due to the compact nature of the $p_z$ orbitals in comparison to the distance between carbon atoms.

For inter-sublattice overlap charge, the two $p_z$ orbitals are separated by the carbon-carbon distance $\vec{R}_{AB}$. Consequently, the leading approximation is given by the following:
\begin{align}
\rho_{\tau_i \bar{\sigma},\tau_j\sigma}(\vec{r})\equiv v^{*}_{\tau_i \bar{\sigma}}(\vec{r})v_{\tau_j\sigma}(\vec{r}) &= \frac{A_0}{N}\sum_{\vec{R}_1,\vec{R}_2}e^{-i\vec{\tau}_i\cdot (\vec{R}_1+\vec{R}_{\bar{\sigma}})+i\vec{\tau}_j\cdot(\vec{R}_2+\vec{R}_{\sigma})}\phi(\vec{r}-\vec{R}_1-\vec{R}_{\bar{\sigma}})\phi(\vec{r}-\vec{R}_2-\vec{R}_{\sigma}) \\ \notag
&\approx A_0 \sum_{\vec{R}}e^{-i(\vec{\tau}_i-\vec{\tau}_j)\cdot\vec{R}} e^{-i\tau_i\cdot\vec{R}_{\bar{\sigma}}+i\tau_j\cdot\vec{R}_{\sigma}}\left( \phi(\vec{r}-\vec{R}-\vec{R}_{\bar{\sigma}})\phi(\vec{r}-\vec{R}-\vec{R}_{\sigma}) \right.\\ \notag
&+ \phi(\vec{r}-\vec{R}-\vec{a}_1-\vec{R}_{\bar{\sigma}})\phi(\vec{r}-\vec{R}-\vec{R}_{\sigma})e^{-i\vec{\tau}_i\cdot\vec{a}_1} +\phi(\vec{r}-\vec{R}-\vec{a}_2-\vec{R}_{\bar{\sigma}})\phi(\vec{r}-\vec{R}-\vec{R}_{\sigma})e^{-i\vec{\tau}_i\cdot\vec{a}_2} \bigg)
\end{align}
We have also numerically verified that the contribution from the next order in the $R$-shell is smaller by an order of magnitude. After substituting the expressions for intra- and intersublattice overlap charges into Eq.\eqref{eq:g0_v1}, we have two independent summations over lattice vectors, resulting from the squaring of the overlap charge. We then apply a change of variables to the center-of-mass coordinates, denoted as $\vec{R}_{cm}$, and relative coordinates, $\vec{R}$. The summation over $\vec{R}_{cm}$ efficiently replaced the integration over the unit cell with an integration across the entire spatial domain. Consequently, we are primarily left with the task of dealing with the summation over $\vec{R}$:


%

\begin{align}
    g_{zz}=&8\sum_{\vec{R}}\left[ \frac{e^2A_0}{4}\int\frac{d^3\vec{q}}{(2\pi)^3}\frac{4\pi |\rho^a(\vec{q})|^2 }{\vec{q}^2}(1-e^{i\vec{q}\cdot\vec{R}_{AB}})e^{i\vec{q}\cdot \vec{R}} \right],\label{eq:gzz} \\
    g_{\perp \perp}=&4\sum_{\vec{R}}\left[\frac{e^2A_0}{4}\int\frac{d^3\vec{q}}{(2\pi)^3}\frac{4\pi |\rho^a(\vec{q})|^2 }{\vec{q}^2}e^{i\vec{q}\cdot \vec{R}} e^{i\vec{R}\cdot \Delta\vec{K}}\right],\label{eq:gpp}\\
    g_{z\perp}=&8\sum_{\vec{R}}
    \left[ \frac{e^2A_0}{4}\int\frac{d^3\vec{q}}{(2\pi)^3}\frac{4\pi |\rho^e(\vec{q})|^2 }{\vec{q}^2}e^{i\vec{q}\cdot \vec{R}}\right],\label{eq:gzp}\\
    g_{\perp z}=&8\sum_{\vec{R}}\left[\frac{e^2A_0}{4}\int\frac{d^3\vec{q}}{(2\pi)^3}\frac{4\pi |\rho^e(\vec{q})|^2 }{\vec{q}^2}e^{i\vec{q}\cdot \vec{R}}\right],\label{eq:gpz} \\
    \rho^a(\vec{q})&=\int d^3\vec{r} \phi(\vec{r})\phi(\vec{r}) e^{-i \vec{q} \cdot \vec{r}} \label{eq:rho_a}\\
    \rho^e(\vec{q})&= \int d^3\vec{r}\bigg( \phi(\vec{r})\phi(\vec{r}-\vec{R}_{ab})  
+e^{i\vec{K}\cdot\vec{a}_2}\phi(\vec{r})\phi(\vec{r}-\vec{a}_2-\vec{R}_{ab})
+e^{i\vec{K}\cdot\vec{a}_1} \phi(\vec{r})\phi(\vec{r}-\vec{a}_1-\vec{R}_{ab})  \bigg)e^{-i \vec{q} \cdot \vec{r}} \label{eq:rho_e}.
\end{align}

\begin{figure}[t]
    \centering \includegraphics[width=1.0\textwidth]{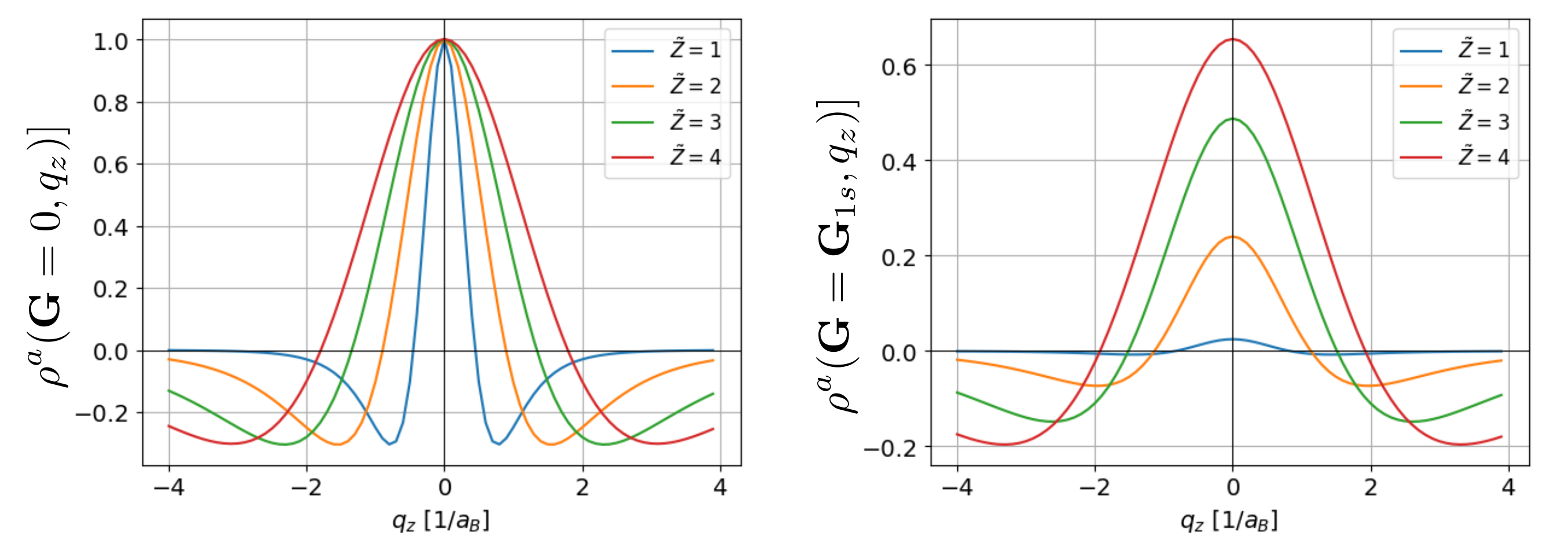}
    \caption{The intrasublattice overlap charge $\rho^a(\vec{G},q_z)$ versus $q_z$. As the effective nuclear charge  $\tilde{Z}$ increases, leading to a more localized wave-function, the Fourier components decay slower.}
    \label{fig:real rho_e}
\end{figure}
\begin{figure}[t]
    \centering \includegraphics[width=1.0\textwidth]{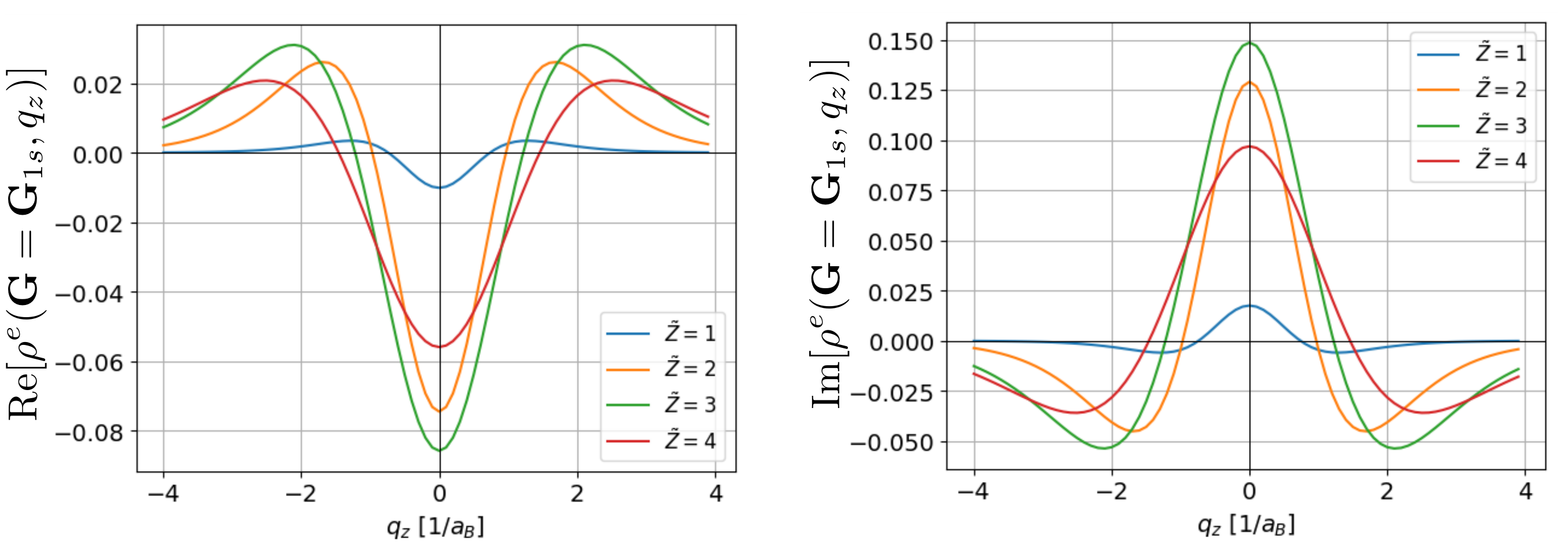}
    \caption{The intersublattice overlap charge $\rho^e(\vec{G}=\vec{G_{1s}},q_z)$ versus $q_z$ for $\vec{G}$. Note the imaginary part is an order of magnitude stronger than the real part.}
    \label{fig:real rho_e}
\end{figure}


Note the terms inside the square brackets $[...]$ decay very slowly as $\vec{R}$ increases. For example, 
\begin{equation}
\int\frac{d^3\vec{q}}{(2\pi)^3}\frac{ |\rho^a(\vec{q})|^2 }{\vec{q}^2}e^{i\vec{q}\cdot \vec{R}} = \int d^3\vec{r}_1 d^3\vec{r}_2 \frac{\phi^2(\vec{r}_1)
     \phi^2(\vec{r}_2)}{|\vec{r}_1-\vec{r}_2+\vec{R}|}\xrightarrow[]{\vec{R}\rightarrow \infty}{}\frac{1}{|\vec{R}|}
\end{equation}
Therefore, given this poor convergence, it becomes more pragmatic to evaluate them in the Fourier space. Using $\sum_{\vec{R}} e^{i\vec{q}\cdot \vec{R}} = \frac{(2\pi)^2}{A_0}\sum_{\vec{G}}\delta^{(2)}(\vec{q}-\vec{G})$ where $\vec{G}$ is the 2D reciprocal lattice vectors, we arrive at the following expressions:
\begin{align}
     &g^{0}_{\perp \perp}=4\sum_{\vec{G}}\frac{e^2}{4}\int \frac{dq_z}{2\pi}\frac{4\pi |\rho^a(\vec{G}- \Delta\vec{K},q_z)|^2}{(\vec{G}- \Delta\vec{K})^2+q_z^2}\label{eq:g_fin_1} \\
     & g^{0}_{zz}=8\sum_{\vec{G}}\frac{e^2 }{4}\int \frac{dq_z}{2\pi}\frac{4\pi |\rho^a(\vec{G},q_z)|^2 }{\vec{G}^2+q_z^2}(1-\cos(\vec{G}\cdot \vec{R}_{ab}))  \\
    &g_{z\perp}^0=8\sum_{\vec{G}}\frac{e^2}{4}\int \frac{dq_z}{2\pi}\frac{4\pi |\rho^e(\vec{G},q_z)|^2}{\vec{G}^2+q_z^2}\\
    &g_{\perp z}^0=8\sum_{\vec{G}}\frac{e^2}{4}\int\frac{dq_z}{2\pi}\frac{4\pi|\rho^e(\vec{G}-\Delta\vec{K},q_z)|^2}{(\vec{G}-\Delta\vec{K})^2+q_z^2} \label{eq:g_fin_2}
\end{align}
Here $\Delta\vec{K}=\vec{K}-\vec{K}^\prime$ is the momentum transfer. From the above equations, it is evident  that they are all positive definite. 
Specifically, the positivity of $g_{\perp\perp}^0$ and $g_{z\perp}^0$ originates from the properties of the exchange integral, as discussed Ref.~\cite{herring1966magnetism}.
We note that the estimates presented in Ref.~\cite{knothe2020quartet} seem to be inaccurate, especially regarding their result indicating $g_{\perp \perp}^0<0$. This discrepancy might stem from their approach of performing the summation in real space instead of in reciprocal space.
Fig.~S2 and Fig.~S3 show the Fourier components of the overlap charge vs $q_z$ for different effective nuclear charge $\tilde{Z}$. Note that $\rho^a$ is real while $\rho^e$ is complex. 



Next, we introduce truncation in $\vec{G}$. For $g^{0}_{\perp \perp}$ and $g^0_{\perp z}$, which involve inter-valley scattering with a momentum transfer of $\Delta \vec{K}$, the primary contributions to the reciprocal lattice sum arise mainly from the $\vec{G}=0$ and the two corners in the first shell closest to $\Delta\vec{K}$, as indicated by the dashed lines in Fig.~(\ref{fig:brillouin zone}).
In contrast, for $g^{0}_{zz}$ and $g^0_{z\perp}$, corresponding to zero-momentum forward scattering, the major contributions arise from the six corners in the first shell, defined as $\vec{G}_{1s}$ in Fig.~(\ref{fig:brillouin zone}). For $g^{0}_{zz}$, the absence of contribution from $\vec{G}=0$ can be attribute to the $1-\cos(\vec{G}\cdot \vec{R}_{ab})$ factor, a result stemming from AA-scattering minus AB-scattering. For $g^{0}_{z\perp}$ 
 the absence of contribution from  $\vec{G}=0$ is consequence of $C_{3}$ rotation symmetry, which implies $\rho^e(0)=0$. 
 Numerical verification has confirmed that contributions from the next shell are significantly smaller, by an order of magnitude, compared to those from the first shell. Consequently, evaluating the overlap charge using the leading  contributions in $\vec{R}'$ and computing $g^0_{\alpha \mu}$ to leading contributions in $\vec{G}$  yield the results summarized in the table below:

\begin{figure}[t]
    \centering
    \includegraphics[width=0.5\textwidth]{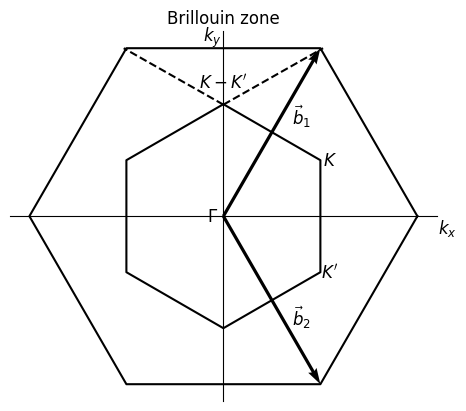}
    \caption{The Brillouin zone of monolayer graphene. Here $\vec{b}_1=\frac{2\pi}{3a}(1,\sqrt{3})$ and $\vec{b}_2=\frac{2\pi}{3a}(1,-\sqrt{3})$, $a \approx 1.42\text{\AA}$ is carbon-carbon distance. The $\Gamma$ point corresponds to $\vec{G}=0$ and the first shell corresponds to $\vec{G}_{1s}=\{G|\text{for $\vec{G}$} \in (\pm \vec{b}_1,\pm \vec{b}_2,\pm(\vec{b}_1+\vec{b}_2))$\}. 
    The two corners in the first shell, which are connected by a dashed line, are equidistant from the zero shell at $\Delta \vec{K}$.}
    \label{fig:brillouin zone}
\end{figure}

%
%

\begin{center}
\begin{tabular}{|c|c|c|c|c|c|c|}
\hline
$meV nm^2$ & $\tilde{Z}=1$ & $\tilde{Z}=1.5$ &$\tilde{Z}=2$ & $\tilde{Z}=2.5$ & $\tilde{Z}=3$& $\tilde{Z}=4$\\
\hline
$g_{\perp \perp}^0$ & 8.88 & 61.5& 156.4& 268.9& 384.3 &600.6\\
\hline
$g_{zz}^0$ & 0.55 &12.55 & 67.46& 184.03 & 353.7 & 781.15 \\
\hline
$g_{z \perp}^0$ &0.234 & 4.41&15.84& 25.11& 25.38 &12.42\\
\hline
$g_{\perp z}^0$ & 5.94 &29.7 & 46.08& 42.93 & 30.42 & 9.72 \\
\hline
\end{tabular}
\end{center}

%
Note that the intersublattice  ( $g_{\perp z}^0$ and $g_{z\perp}^0$) contribution is almost an order of magnitude smaller than the intra-sublattice ($g_{\perp \perp}^0$ and $g_{zz}^0$) contribution. However, the values for $g_{\perp z}^0$ and $g_{z\perp}^0$ actually has a strong contribution coming from electron-optical-phonon scattering. The estimates for these interactions \cite{wu2018theory} are:
\begin{equation}\label{eq:g0_estimate2}
    (g_{\perp z}^{0},\, g_{z\perp}^{0})_{phonon} = (-69, -52)\text{meV} \cdot \text{nm}^2
\end{equation}
By setting the effective nucleus charge $\tilde{Z}=2.5$, a value slightly reduced from the $\tilde{Z}=3$ typically associated with an isolated carbon atom \cite{clementi1963atomic}, and factoring in the electron-phonon contributions, we derived the following estimates:

\begin{equation}\label{eq:g0_estimate}
(g_{\perp \perp}^0,g_{zz}^0 , g_{\perp z}^{0},\, g_{z\perp}^{0})
= (269 ,  184 , -26, -27)\text{meV}\cdot\text{nm}^2.
\end{equation}
It is important to point out that our central result, namely that an increase in $\kappa$ leads to a more pronounced decrease in $u_\perp$ relative to the increase of $u_z$, remains consistent regardless of the choice of $\tilde{Z}$.
To leading order in electron-electron and  electron-phonon contribution, $g_{\perp0}^0=g_{0\perp}^0=g_{z0}^0=g_{0z}^0=0$ \cite{kharitonov2012phase}. 

\section{Calculation Details of Landau level mixing}

In the weakly interacting systems, the perturbative expansion of low-energy effective Hamiltonian reads that $\hat{H}= \hat{H}_0 + \hat{V}^{(1)} + \hat{V}^{(2)} +...$, where $\hat{H}_0$ is the non-interacting Hamiltonian, $\hat{V}^{(1)}$ is the projection of the bare interactions onto the degenerate non-interactng ground state manifold of energy $E_0$. In this study, the non-interacting ground states have two electrons in the empty $n=0$ Landau level (LL) along with fully empty(occupied) $n> 0(n<0)$ LLs \cite{sodemann2013landau}.The next order correction to the effective Hamiltonian is obtained from the second order perturbation theory,
\begin{align}\label{eq:effective_asym}
    \hat{V}^{(2)} &= -\mathcal{P}(\hat{H}_{s}+\hat{H}_{a}) \mathcal{P}_{\perp}\frac{1}{\hat{H}_0 - E_{0}} \mathcal{P}_{\perp}(\hat{H}_{s}+\hat{H}_{a})\mathcal{P} \notag\\
    &= -\mathcal{P}\hat{H}_{s} \mathcal{P}_{\perp}\frac{1}{\hat{H}_0 - E_{0}} \mathcal{P}_{\perp}\hat{H}_{s}\mathcal{P} - \left(\mathcal{P}\hat{H}_{s} \mathcal{P}_{\perp}\frac{1}{\hat{H}_0 - E_{0}} \mathcal{P}_{\perp}\hat{H}_{a}\mathcal{P} + h.c.\right) -\mathcal{P}\hat{H}_{a} \mathcal{P}_{\perp}\frac{1}{\hat{H}_0 - E_{0}} \mathcal{P}_{\perp}\hat{H}_{a}\mathcal{P} \notag\\
    &\equiv \hat{V}_{ss}^{(2)} + \hat{V}_{sa}^{(2)} + \hat{V}_{aa}^{(2)}.
\end{align}
The first term $\hat{V}_{ss}^{(2)}$ has been studied in previous works and could potentially play a crucial role in certain fractional quantum Hall states \cite{peterson2013more,peterson2014effects}. However, due to its spin-valley SU(4) symmetry, it cannot differentiate competing spin-valley orders and will not be further discuss here. $\hat{V}_{sa}$ gives rise to first-order-in-$\kappa$ corrections to the valley-dependent effective interactions, while the quadratic-in-$g_{\alpha\mu}^{0}$ corrections $\hat{V}_{aa}$ is independent of $\kappa$ (\textit{i.e.}, dielectric screening environment).

As we briefly explained in the main text, $\hat{V}^{(2)}$ admits the decomposition $\hat{V}^{(2)}=\hat{V}_{3b}+\hat{V}_{2b}+\hat{V}_{1b}+constant$ based on the number $n_{v}$ of the virtually excited single-particle states. Note that $n_v\neq 0$ since the intermediate many-body states must have energy different from $E_0$. The $n_v=1$ and $2$ terms give rise to three-body interactions $\hat{V}_{3b}$ and two-body interactions $\hat{V}_{2b}$ \cite{sodemann2013landau}, respectively.  
$n_v=3$ corresponds to a SU(4) symmetric single-particle self-energy $\hat{V}_{1b}$ and can be omitted together with the constant energy shift generated by the $n_{v}=4$ term.
$\hat{V}_{3b}=0$ in the zeroth LL of monolayer graphene thanks to the particle-hole symmetric bare Hamiltonian and the particle-hole symmetry between the $N>0$ and $N<0$ remote LLs \cite{peterson2013more}. The corrections to valley-dependent effective interactions are entirely contributed by $\hat{V}_{2b}$.

To study the two-body interactions preserving the translational and rotational symmetries, it is convenient to use center of mass and relative coordinates,
\begin{equation}\label{eq:pp_basis}
    A=\frac{a_{i}+a_{j}}{\sqrt{2}},\ B=\frac{b_{i}+b_{j}}{\sqrt{2}},\ a=\frac{a_{i}-a_{j}}{\sqrt{2}},\ b=\frac{b_{i}-b_{j}}{\sqrt{2}},
\end{equation}
where $a_i,b_i$ are the lowering operators in the LL and guiding center space, respectively. We construct states with well defined center of mass and relative coordinate quantum numbers,
\begin{equation}
    |N M\rangle_{c}|n m\rangle_{r}=\frac{A^{\dagger N} B^{\dagger M} a^{\dagger n} b^{\dagger m}}{\sqrt{N ! M ! n ! m !}}|0\rangle.
\end{equation}
For a two-body interaction $\hat{V}$ preserving the translational and rotational symmetries,
\begin{equation}
    {}_c\bra{N'M'}{}_r\bra{n'm'}\hat{V}\ket{NM}_{c}\ket{nm}_r={}_{r}\bra{n'm'}\hat{V}\ket{nm}_{r}\delta_{m'-n',m-n}\delta_{N',N}\delta_{M',M},
\end{equation}
where the subscript `c' and `r' stands for the center of mass and relative coordinates, respectively. ${}_r\langle 0m| \hat{V}|0m\rangle_r$ are the Haldane pseudopotentials in the zeroth LL. We follow Ref.~\cite{sodemann2013landau} to evaluate in the next two sections the corrections to the valley-dependent Haldane pseudopotentials, $u_{\alpha}^{(2)}(m)/2$ induced by $V_{sa}^{(2)}$ and $u_{\alpha}^{(2)'}(m)/2$ induced by $V_{aa}^{(2)}$.

\subsection{$V_{sa}-$induced Haldane pseudopotential corrections $u_{\alpha}^{(2)}(m)$}

$u_{\alpha}^{(2)}(m)$ receive contributions from three types of non-vanishing Feynman diagrams depicted in Fig.2: (a) the particle-particle ladder diagram, (b) the double-exchange diagram, and (c) the vertex correction to the valley-dependent interactions. As discussed in the main text,
\begin{equation}
    u_{\alpha}^{(2)} (m) = \frac{\kappa }{2\pi l_{B}^2}\sum_{\mu=0,x,y,z}(c_{m,a}^{\mu}+c_{m,b}^{\mu}+c_{m,c}^{\mu})g_{\alpha\mu}^{0}.
\end{equation}
Because $g_{\alpha x}^{0} = g_{\alpha y}^{0} = g_{\alpha\perp}^{0}$, it will be convenient to define $c_{m,a/b/c}^{\perp} = c_{m,a/b/c}^{x} + c_{m,a/b/c}^{y}$.

It is straightforward to derive $c_{m,a}^{0,z}$ under two-particle bases ($c_{m,a}^{\perp}=0$ due to the valley-sublattice locking in the zeroth LL, see the main text.):
\begin{equation}\label{eq:c_ma}
    \frac{c_{m,a}^{0,z}}{4\pi}= -2\sum_{n=1}^{2N_{c}}\prescript{}{r}{\bra{0m}}\delta(\bm r_{ij})\ket{n\ n+m}_{r}\prescript{}{r}{\bra{n\ n+m}}\frac{1}{r_{ij}}\ket{0m}_{r}\sum_{n_{i,j}=-\infty}^{\infty}\frac{\Theta(n_in_j)}{|\epsilon_{\overline{n_i}}+\epsilon_{\overline{n_j}}|}|\prescript{}{c}\langle 0|\prescript{}{r}\langle n||n_{i}|,|n_{j}|\rangle|^2.
\end{equation}
We measure the length in the unit of $l_B$ and the kinetic energy in the unit of $\hbar v_F/l_B$ in this section. The interaction matrix elements are derived as follows:
\begin{align}
    {}_r\bra{n(m+n)}\frac{1}{r_{ij}}\ket{0m}_{r}&=\int\frac{d^2 q}{(2\pi)^2}\frac{2\pi}{|\bm q|}\mathcal{F}_{n,0}(\sqrt{2}\bm q)\mathcal{F}_{m+n,m}^{*}(-\sqrt{2}\bm q) 
    \label{eq:symele1}\\
    &=\frac{(-1)^n}{2}\sqrt{\frac{(n+m)!}{n!m!}}\frac{\Gamma(n+\frac{1}{2})}{n!}{}_2F_{1}(-m,n+\frac{1}{2};n+1;1)\label{eq:symele2},
\end{align}
\begin{align}\label{eq:asymele}
    {}_r\bra{n(m+n)}\delta(\bm r_{ij})\ket{0m}_{r}&=\int\frac{d^2 q}{(2\pi)^2}\mathcal{F}_{n,0}(\sqrt{2}\bm q)\mathcal{F}_{m+n,m}^{*}(-\sqrt{2}\bm q) \notag\\
    &=\frac{(-1)^n}{4\pi}\sqrt{\frac{(n+m)!}{n!m!}}{}_2F_{1}(-m,n+1;n+1;1)=\frac{(-1)^n}{4\pi}\delta_{m,0}.
\end{align}
In Eq.~\eqref{eq:symele1}, we used the density form factor $\mathcal{F}_{n',n}(\bm q)=\bra{n'}e^{i\bm q\cdot(\hat{z}\cross\bm\Pi)}\ket{n}$ of the LLs of the two dimensional electron gas and an analog relation for the guiding center coordinates $\bra{m'}e^{i\bm q\cdot\bm R}\ket{m}=\mathcal{F}_{m',m}^{*}(-\bm q)$. $\mathcal{F}_{n',n}(\bm q)=\mathcal{F}_{n,n'}^{*}(-\bm q)$ and for $n'\geq n$, 
\begin{equation}\label{eq:f}
    \mathcal{F}_{n',n}(\bm q)=\sqrt{\frac{n !}{n^{\prime} !}}\left(\frac{q_{x}+i q_{y}}{\sqrt{2}}\right)^{n^{\prime}-n}  L_{n}^{n^{\prime}-n}\left(\frac{|q|^{2}}{2}\right)e^{-|q|^{2} / 4},
\end{equation}
with $L_n^{m}(x)$ the generalized Laguerre polynomial. The $\sqrt{2}$ factor in Eq.~\eqref{eq:symele1} is because we are using the relative coordinates Eq.~\eqref{eq:pp_basis}. ${}_2F_{1}$ is the hypergeometric function. Eq.~\eqref{eq:asymele} indicates that a pair of electrons with nonzero relative angular momentum do not interact via contact interaction because their wave functions vanish at $\bm r_{ij}=0$.

The last term in Eq.~\eqref{eq:c_ma} is obtained from the relation
\begin{equation}
\left|n_{1},n_{2} \right\rangle= \sum_{n=0}^{n_{1}+n_{2}} R_{n_{2}, n}^{n_{1}+n_{2}}\times\left|n_{1}+n_{2}-n\right\rangle_{c}|n\rangle_{r},\quad (n_{1,2}\geq 0)
\end{equation}
where $R^{L}$ is a $L\times L$ real, symmetric and orthogonal matrix whose explicit form reads that
\begin{equation}\label{eq:R}
    R_{n',n}^{L}=\sqrt{\frac{\binom{L}{n'}}{2^{L}\binom{L}{n}}}\sum_{\nu=\text{max}(0,n+n'-L)}^{\text{min}(n,n')}\binom{L-n'}{n-\nu}\binom{n'}{\nu}(-1)^{\nu}.
\end{equation}
Only $R_{n',n}^{n}=(-1)^{n}\sqrt{\binom{n}{n'}/2^{n}}$ is needed in Eq.~\eqref{eq:c_ma} because the center of mass kinetic quantum number $n_1+n_2-n =0$. Combining Eq.~\eqref{eq:c_ma}-\eqref{eq:R}, we obtain $c_{m,a}^{0,z} = 0$ for $m\neq 0$ and as shown in Eq.(9) in the main text,
\begin{equation}
        \frac{c_{m=0,a}^{0,z}}{4\pi}= \sum_{n=1}^{2N_{c}}\frac{\Gamma(n+\frac{1}{2})}{2^{n+1} n!}\sum_{n_1=0}^{n}\frac{\binom{n}{n_{1}}}{\sqrt{2n_1}+\sqrt{2(n-n_1)}}. \label{eq:cm_pp}
\end{equation}
Here the cut off of the LL numbers is $2N_c$ because $n=|n_i|+|n_j|$ and $|n_{i,j}|\leq N_c$.

The expressions for $c_{m,b/c}^{\mu}$ are given by Eqs.(10)-(11) in the main text. 
In the interaction matrix elements $\mathcal{V}$ and $\mathcal{V}^{\mu}$, we introduced the graphene form factors $F_{\xi'n',\xi n}^{\mu}(\bm q)=\bra{\overline{\xi'n'}}\sigma^{\mu} e^{i\bm q\cdot(\hat{z}\times\bm\Pi)}\ket{\overline{\xi n}}$ for $\mu=0,\pm,z$ \footnote{We take the convention $\sigma^{\pm} = (\sigma^{x}\pm i\sigma^{y})/2$},
\begin{align}
    &F_{\xi' n', \xi n}^{0/z}(\bm q)= \frac{1}{2^{1-(\delta_{\xi,0}+\delta_{\xi',0})/2}}\left(\mathcal{F}_{n',n}(\bm q) \pm \xi\xi'\mathcal{F}_{n'-1,n-1}(\bm q)\right),\label{eq:f0z}\\
    &F_{\xi' n', \xi n}^{+}(\bm q)= \frac{1}{2^{1-(\delta_{\xi,0}+\delta_{\xi',0})/2}}\xi\mathcal{F}_{n',n-1}(\bm q), \label{eq:f+}\\
    &F_{\xi' n', \xi n}^{-}(\bm q)= \frac{1}{2^{1-(\delta_{\xi,0}+\delta_{\xi',0})/2}}\xi'\mathcal{F}_{n'-1,n}(\bm q), \label{eq:f-}
\end{align}
where $\xi,\xi'=\pm 1$ and $n,n'\geq 0$. Here we require that the density form factor $\mathcal{F}_{n',n}(\bm q)=0$ if $n'<0$ or $n<0$. Using the relation $\prescript{}{r}{\bra{m}}e^{i\bm q\cdot\bm \eta_{ij}}\ket{m}_r = e^{-\frac{\bm q^2}{2}} L_{m}(\bm q^2)$, we find that
\begin{align}
    \frac{c_{m,b}^{z}}{4\pi} &=\iint\frac{d^2\bm q_{1}d^2\bm q_{2}}{(2\pi)^4}\frac{2\pi}{|\bm q_2|}e^{-\bm q_1^{2}-\bm q_2^2+\bm q_1\cdot\bm q_2}L_{m}(|\bm q_1-\bm q_2|^2)\sideset{}{'}\sum_{n,n'=0}^{N_c}\frac{1}{\sqrt{2n}+\sqrt{2n'}}\frac{1}{n!n'!}\left(\frac{\bar{q}_1 \bar{q}_2^{*}}{2}\right)^{n}\left(\frac{\bar{q}_1^{*}\bar{q}_2}{2}\right)^{n'} \notag\\
    &=\int_{0}^{\infty}\int_{0}^{\infty} \frac{dq_1dq_2}{(2\pi)^2}\int_{0}^{2\pi} d\theta\ q_1e^{-q_1^2-q_2^2-q_1q_2\cos\theta}L_{m}(q_1^2+q_2^2-2q_1q_2\cos\theta)\sideset{}{'}\sum_{n,n'=0}^{N_c}\frac{1}{\sqrt{2n}+\sqrt{2n'}}\frac{1}{n!n'!}\left(\frac{q_1 q_2}{2}\right)^{n+n'}e^{i(n'-n)\theta}, \label{eq:c_mb^0z}
\end{align}
\begin{align}
    \frac{c_{m,b}^{\perp}}{4\pi} &=-2\iint\frac{d^2\bm q_{1}d^2\bm q_{2}}{(2\pi)^4}\frac{2\pi}{|\bm q_2|}e^{-\bm q_1^{2}-\bm q_2^2+\bm q_1\cdot\bm q_2}L_{m}(|\bm q_1-\bm q_2|^2)\sum_{n,n'=1}^{N_c}\frac{1}{\sqrt{2n}+\sqrt{2n'}}\frac{\sqrt{nn'}}{n!n'!}\left(\frac{\bar{q}_1 \bar{q}_2^{*}}{2}\right)^{n}\left(\frac{\bar{q}_1^{*}\bar{q}_2}{2}\right)^{n'}\left(\frac{\bm q_1^2}{2}\right)^{-1} \notag\\
    &=-4\int_{0}^{\infty}\int_{0}^{\infty} \frac{dq_1dq_2}{(2\pi)^2}\int_{0}^{2\pi} d\theta\ \frac{1}{q_1}e^{-q_1^2-q_2^2-q_1q_2\cos\theta}L_{m}(q_1^2+q_2^2-2q_1q_2\cos\theta)\sum_{n,n'=1}^{N_c}\frac{1}{\sqrt{2n}+\sqrt{2n'}}\frac{\sqrt{nn'}}{n!n'!}\left(\frac{q_1 q_2}{2}\right)^{n+n'}e^{i(n'-n)\theta}, \label{eq:c_mb^p}
\end{align}
with $\bar{q}_i\equiv q_{i,x} + iq_{i,y}$.
\begin{align}
    c_{m,c}^{0/z} =-\text{Re}&\int_{0}^{\infty}\int_{0}^{\infty} \frac{dq_1dq_2}{(2\pi)^2}\int_{0}^{2\pi} d\theta\ q_1 e^{-q_1^2-\frac{q_2^2}{2}-i q_1q_2\sin\theta}L_{m}( q_1^2)\left[\sideset{}{'}\sum_{n'\geq n\geq 0}\frac{1}{\sqrt{2n}+\sqrt{2n'}}\frac{1}{2^{n'}n'!}q_1^{n'-n}q_2^{n+n'}e^{i(n'-n)\theta}\right.\notag\\
    &\times \left.\left(L_{n}^{n'-n}(\frac{q_{1}^2}{2}) \mp L_{n-1}^{n'-n}(\frac{q_{1}^2}{2})\right) + \sum_{n> n'\geq 0}\frac{1}{\sqrt{2n}+\sqrt{2n'}}\frac{1}{2^{n}n!}(-q_1)^{n-n'}q_2^{n+n'}e^{i(n'-n)\theta}\left(L_{n'}^{n-n'}(\frac{q_{1}^2}{2}) \mp L_{n'-1}^{n-n'}(\frac{q_{1}^2}{2})\right)\right] \notag\\
    =-\text{Re}&\int_{0}^{\infty}\frac{dq_1}{2\pi}q_1e^{-q_1^2}L_{m}( q_1^2)\sideset{}{'}\sum_{n,n'=0}^{N_{c}}\frac{1}{\sqrt{2n}+\sqrt{2n'}}\frac{1}{2^{l}l!}q_1^{|n-n'|}I_{n,n'}(q_1)\left(L_{j}^{|n-n'|}(\frac{q_{1}^2}{2}) \mp L_{j-1}^{|n-n'|}(\frac{q_{1}^2}{2})\right). \label{eq:c_mc^0z}
\end{align}
In the last line, we defined $j=\min(n,n')$, $l=\max(n,n')$, and the integral \cite{gradshteyn2014table}
\begin{equation}
    I_{n,n'}(q_1) = \int_{0}^{\infty} dq_2 q_2^{n+n'} J_{|n-n'|}(q_1q_2)e^{-\frac{q_2^2}{2}} = \frac{2^{l+\frac{1}{2}} \Gamma \left(l+\frac{1}{2}\right)}{2^{|n-n'|+1}\Gamma(|n-n'|+1)}
\, _1F_1\left(l+\frac{1}{2};|n-n'|+1;-\frac{q_1^2}{2}\right)  q_1^{|n-n'|}.
\end{equation}
Here, $\ _{1}F_{1}$ is a confluent hypergeometric function. The summations in Eq.~\eqref{eq:c_mb^0z} and~\eqref{eq:c_mc^0z} exclude the $n=n'=0$ case. Our numerical results of Eqs.~\eqref{eq:c_mb^0z}-\eqref{eq:c_mc^0z}, summarized in Tables I(a) and (b) in the main text, indicate that $c_{0,b}^{\mu}$and $c_{0,c}^{z}$ logarithmically diverge with respect to the LL cutoff $N_c$ as $c_{0,a}^{0/z}$ do, whereas $c_{0,c}^{0}$ and $c_{m\neq 0, b/c}^{\mu}$ converge.

\begin{figure}
    \centering
    \includegraphics[width=0.4\linewidth]{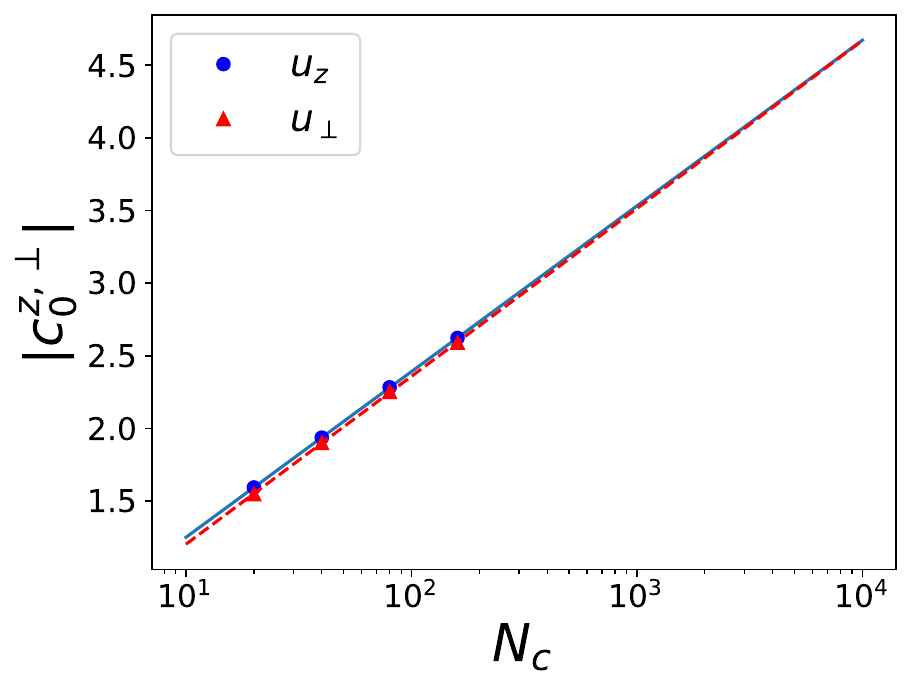}
    \caption{The logarithmic dependence of $|c_0^{z,\perp}|$ on the LL cutoff $N_c$. The blue dots and red triangles represent numerical results for $c_{0}^{z}$ and $-c_{0}^{\perp}$, respectively. The dashed lines are linear fit in the semi-log plot, which indicate $|c_0^{z,\perp}|\approx 0.11 + 0.5 \ln{N_c}$. }
    \label{fig:c}
\end{figure}

The logarithmic divergence of the contact valley-dependent interactions in graphene have been studied previously in the zero-field with the renormalization group (RG) analysis \cite{kharitonov2012phase}. It was shown that the sublattice-and valley-dependent short-range interactions obey the RG flow equations $d\bar{g}_{\alpha 0}/d\xi=0$ and $d\bar{g}_{\alpha z}/d\xi = 4F(w)(\bar{g}_{\alpha z}-\bar{g}_{\alpha \perp})$, where $F(w)\approx \kappa/4$ for small $\kappa$, $\xi=\ln{l/a_0}$ and $l$ is the running length scale. Ref.~\cite{kharitonov2012phase} then estimates the effective valley-dependent interactions in the zeroth LL in the magnetic field as $2\pi l_{B}^2u_{\alpha} = \bar{g}_{\alpha 0}(l=l_B) + \bar{g}_{\alpha z}(l=l_B)$. In the limit $\kappa\ln{l_B/a_0}\ll 1$, the RG equation yields
\begin{align}
    2\pi l_{B}^2u_{\alpha} &\approx g_{\alpha 0} +g_{\alpha z}(1 + \kappa \ln\frac{l_{B}}{a_0}) - \kappa g_{\alpha \perp}\ln\frac{l_{B}}{a_0} \label{eq:kharitonov}\\
     &= g_{\alpha 0} +g_{\alpha z}(1 + 0.5\kappa \ln{N_{c}}) - 0.5\kappa g_{\alpha \perp}\ln{N_{c}} + const., 
\end{align}
To arrive at the second line, we note that $N_c$ depends on the out-of-plane magnetic field such that $\sqrt{N_c}/l_B\propto$ a fixed energy cutoff $\Lambda$.
The coefficient of the first-order-in-$\kappa\ln{N_c}$ term agrees with our results, see Fig.~\ref{fig:c}. 
We note that the ratio $u_{z}/u_{\perp}$ is affected by both $\kappa$ and $l_B$. Thus, there can be a quantum phase transition between the KD and AF states induced by either by out-of-plane magnetic field (via $l_B$) or the dielectric screening environment (via $\kappa$). In the main text, we primarily explore the transition induced by changes in the dielectric environment. However, it's important to mention that a field-induced phase transition also occurs for a moderate fine structure constant ($\kappa=0.5\sim 1$). Specifically, in a strong field scenario ($N_c=100$), the system resides in the AF state, whereas in a weak field ($N_c=7000$), it transitions to the KD state.
Due to the logarithmic dependence of $u_{z}$ and $u_{\perp}$ on the magnetic field, adjusting the dielectric environment is a more effective means of modulating the phase boundary than varying the magnetic field strength.

Here we have taken a common UV cutoff for all the short-range interactions comparable to the lattice constant. This is appropriate for those that originate from the Coulomb forces. For those arising from optical phonons it would be more appropriate to take the UV cutoff to be comparable to the phonon gap. However, since the phonon gap is smaller than the band-width of the $p_z$ orbitals,  accounting for this would mean that the contribution to $u_{\alpha}$ coefficients from the interactions mediated by phonons (i.e. $g_{\perp z}^{0}, g_{z\perp}^{0}$ as discussed in Eq.\eqref{eq:g0_estimate2}), would have a smaller enhancement factor due to renormalizations than the one we have used here, because the phonon interactions would effectively renormalize over a shorter RG time. However, since the phonon contribution to interaction coefficients is already smaller than the Coulomb contribution, accounting for this would not significantly alter our estimates. For example, this would mean that the  coefficient $c_{0}^{z}$ that accompanies the $K$-optical-phonon
potential $g_{\perp z}$ contribution to $u_{\perp}$ would be smaller than the one listed before (e.g. in Eq.\eqref{eq:kharitonov}), because it would have smaller UV cutoff. However, this contribution to the linear in $\kappa$ enhancement of $u_{\perp}$ is smaller compared to that coming from the coefficient $g^0_{\perp \perp}$ generated by the Coulomb interactions. Thus we expect accounting for this would not significantly alter our estimates.

\subsection{$\hat{V}_{aa}-$ induced Haldane pseudopotential corrections to  $u_{\alpha}(m)$}

\begin{figure}
    \centering
    \includegraphics[width=0.75\linewidth]{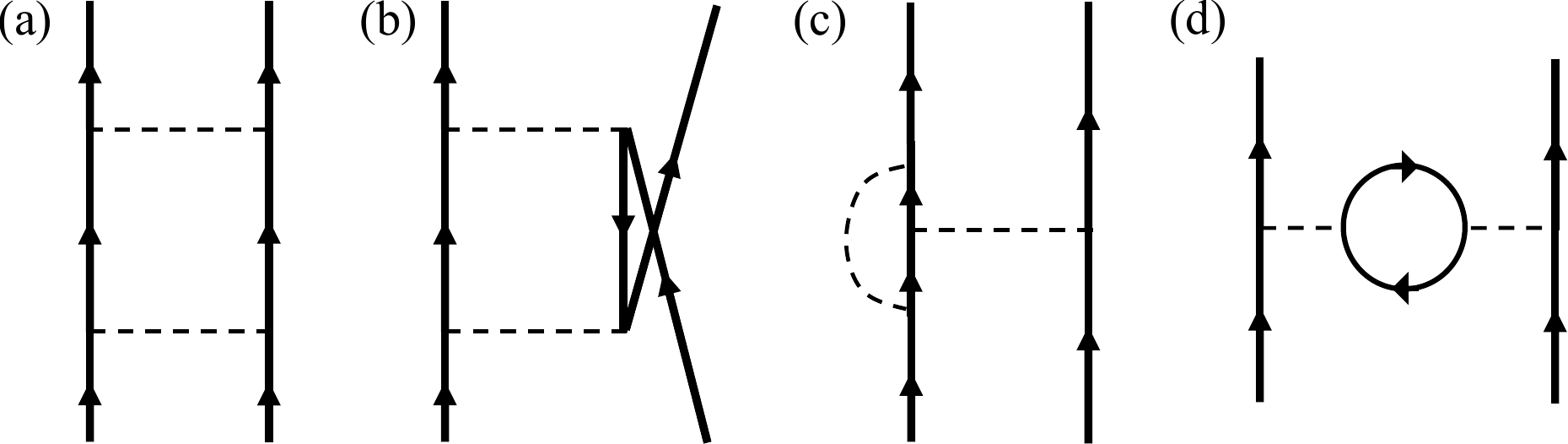}
    \caption{Feynman diagrams representing the quadratic-in-$g_{\alpha\mu}^{0}$ corrections to two-body valley-dependent Haldane pseudopotentials in the zeroth LL. The dashed lines stand for the contact sublattice-and valley-dependent interactions.}
    \label{fig:feynman_diagram2}
\end{figure}

The non-vanishing valley-dependent interactions generated $\hat{V}_{aa}$ can be represented by Feynman diagrams depicted in Fig.~\ref{fig:feynman_diagram2}a (particle-particle), b (double-exchange), c (vertex correction), and d (bubble).  The first three types of Feynman diagrams can be evaluated according to the same procedure outlined in the previous section and we will explain later the details to compute the bubble diagrams, which cannot be generated by $\hat{V}_{sa}$. The results show that the main contributions from $\hat{V}_{aa}$ are the corrections to $m=0$ Haldane pseudopotentials that diverge as $\sqrt{N_c}$,
\begin{align}
   u_{\perp}^{(2)'}(m=0)=&\frac{1}{2\pi l_{B}^2\bar{g}}\left[g_{zz}^{0}g_{\perp z}^{0}z_a-g_{zz}^{0}g_{\perp z}^{0}z_b+2(g_{z z}^{0}g_{\perp\perp}^{0}+g_{z\perp}^{0}g_{\perp z}^{0})y_b-4g_{z\perp}^{0}g_{\perp\perp}^{0}x_b - g_{zz}^{0}g_{\perp z}^{0}z_c-2g_{z\perp}^{0}g_{\perp z}^{0}y_c - (g_{\perp z}^{0})^2z_d\right], \label{eq:2nd-in-g1}\\
   u_{z}^{(2)'}(m=0)=&\frac{1}{2\pi l_{B}^2\bar{g}}\left[(g_{\perp z}^{0})^2z_a-(g_{\perp z}^{0})^2z_b+4g_{\perp z}^{0}g_{\perp\perp}^{0}y_b-4(g_{\perp\perp}^{0})^2x_b+ g_{zz}^{0}(g_{zz}^{0}-2g_{\perp z}^{0})z_c+2g_{zz}^{0}(g_{z\perp}^{0}-2g_{\perp\perp}^{0})y_c- (g_{z z}^{0})^2z_d\right], \label{eq:2nd-in-g2}
\end{align}
where we have kept $g_{0\mu}=g_{\alpha 0}=0$ for simplicity, $\bar{g}=\hbar v_F l_B / \sqrt{N_c}$, and the coefficients $z_a\approx 0.11, z_b\approx y_b\approx x_b\approx 0.1, z_{c}=0.2, y_c=0.017,$ and $z_d=0.45$. The subscripts of these coefficients represent the associated Feynman diagrams. For the bubble diagram,
\begin{align}
    \frac{u_{\alpha,d}^{(2)\prime}(m)}{2} &= -4\times\frac{\left(g_{\alpha z}^{0}\right)^2}{2\pi l_B^2} \int\frac{d^2q}{(2\pi)^2}\prescript{}{r}{\langle m|}e^{i\bm{q}\cdot\bm{\eta}_{ij}}|m\rangle_{r}\sideset{}{'}\sum_{n,n'=-\infty}^{\infty}\frac{\Theta(\epsilon_{n})-\Theta(\epsilon_{n'})}{\epsilon_{n}-\epsilon_{n'}}|F_{n',n}^{z}(\bm q)F_{0,0}^{z}(\bm q)|^2 \notag\\
    &= -\frac{1}{2\pi l_B^2} \frac{2\left(g_{\alpha z}^{0}\right)^2}{\hbar v_F l_B} \int\frac{d^2\bm q}{(2\pi)^2}e^{-\frac{3\bm q^2}{2}}L_{m}(\bm q^2)\sideset{}{'}\sum_{n,n'=0}^{N_c}\frac{1}{\sqrt{2|n|}+\sqrt{2|n'|}}\frac{j!}{l!}\left(\frac{\bm q^2}{2}\right)^{l-j}\left(L_{j}^{l-j}(\frac{\bm q^2}{2})+L_{j-1}^{l-j}(\frac{\bm q^2}{2})\right)^2.
\end{align}
where $j=\min(n,n')$, $l=\max(n,n')$, and the summation excludes the $n=n'=0$ case. Our numerical calculations suggest that $u_{\alpha,d}^{(2)'}(m)\approx -z_d\frac{\left(g_{\alpha z}^{0}\right)^2}{2\pi l_B^2\bar{g}}$ for large $N_c$ with $z_d$ converging to $\sim 0.45$.
$u_{\alpha,d}^{(2)'}(m>0)$ is at least one order of magnitude smaller than $u_{\alpha,d}^{(2)'}(m=0)$ and thus the valley-dependent interactions remains short-ranged, although we would like to mention that $|u_{\alpha,d}^{(2)'}|\gg |u_{\alpha,a/b/c}^{(2)'}|$ for $m>0$, indicating that the effective interactions generated by the bubble diagrams can have a longer interaction range than the other three channels.

We comment that the square root divergence of $\delta{u}_{\perp,z}'(0)$ causes significant uncertainties on the estimation of their values. For instance, if we fix $g_v N_c/2\pi l_{B}^2=$number of graphene unit cell per area ($g_v=2$ is the valley degeneracy), $\bar{g}=\hbar v_{F}a_0\sqrt{\sqrt{3}/2\pi}\approx 85\textrm{meV}\cdot\textrm{nm}^2$ and, according to the estimated strength of bare valley-dependent interactions, Eq.~\eqref{eq:g0_estimate}, $2\pi l_B^2 u_{\perp}^{(2)\prime}(m=0)\sim 159\textrm{meV}\cdot\textrm{nm}^2$ and $2\pi l_B^2 u_z^{(2)\prime}(m=0)\sim -492\textrm{meV}\cdot\textrm{nm}^2$. On the other hand, if $\hbar v_F \sqrt{2N_c}/l_B =\Lambda \sim 2$eV, the energy cut off of the continuum model of graphene, $2\pi l_B^2 u_{\perp}^{(2)\prime}(m=0)\sim 44\textrm{meV}\cdot\textrm{nm}^2$ and $2\pi l_B^2 u_z^{(2)'}(m=0)\sim -137\textrm{meV}\cdot\textrm{nm}^2$. Despite these uncertainties on the estimates, we can see that $u_{\alpha}^{(2)'}(m=0)$ are overall smaller than $u_{\alpha}^{(2)}(m=0)$ depicted in Fig.1 for large $\kappa>0.5$, although their ratio is much larger than the ratio of the bare lattice-scale interaction and long-range Coulomb interaction, $\mathcal{O}(g_{\alpha\mu}^{0}\epsilon /e^2 l_B)$, due to the $\sqrt{N_c}$ divergence in Eqs.~\eqref{eq:2nd-in-g1} and~\eqref{eq:2nd-in-g2}. 
More importantly, we emphasize that unlike $u_{\alpha}^{(2)}(m=0)$ generated by $\hat{V}_{sa}^{(2)}$, $u_{\alpha}^{(2)'}(m=0)$ are insensitive to the long-wave dielectric screening (\textit{i.e.}, $\kappa$), which justifies neglecting $u_{\alpha}^{(2)'}(m=0)$ in the study of the dielectric screening effects on the valley-dependent Haldane pseudopotentials in the main text.

\end{document}